# An Improved Scheme for Pre-computed Patterns in Core-based SoC Architecture

Elaheh Sadredini[1,2], Reza Rahimi[3], Paniz Foroutan[2], Mahmood Fathy[1] and Zainalabedin Navabi[2]
el_sadredini@comp.iust.ac.ir, g_rahimi@ce.sharif.edu, mahfathy@iust.ac.ir, {paniz, navabi}@cad.ut.ac.ir

1. *Computer EngineeringDepartment, Iran University of Science and Technology, Iran*
2. *Electrical and Computer Engineering Department, University of Tehran, Iran*
3. *Computer Engineering Department, Sharif University of Technology, Iran*

***Abstract***

*By advances in technology, integrated circuits have come to include more functionality and more complexity in a single chip. Although methods of testing have improved, but the increase in complexity of circuits, keeps testing a challenging problem. Two important challenges in testing of digital circuits are test time and accessing the circuit under test (CUT) for testing. These challenges become even more important in complex system on chip (SoC) zone. This paper presents an improved scheme for generating pre-computed test patterns in core-based systems on chip. This approach reduces the number of pre-computed test patterns and as the result, test application time (TAT) will be decreased. Experimental results on ISCAS'89 benchmark circuits show improvement in the number of test clock cycles.*

## 1. Introduction

As a result of advances in technology of integrated digital circuits both in size and dimension, more complex SoC architectures have evolved [1][2]. Although test and testing strategies have also moderately improved, new challenges introduced by this progress necessitate more improvements in this area [3]. The issue of porting test data to specific cores for complex systems (e.g., SoCs) has become more important due to such improvements. As a system gets larger and more complex, the cost of providing test data for its internal components increases. In addition, the overhead of memory requirement for saving test data is high. Therefore, many works have been done to decrease the number of test vectors which are stored in the memory.

For achieving reduction in test application time (TAT) and ATE memory requirements, many approaches try to compress tests by using pattern overlapping [4-6]. The work presented in [7] uses bitmasks to provide significant improvement in compression. The authors in [10] improve the work [7] by suggesting a slice partitioning along with a multiple dictionaries bitmask approach, and also a slice bit reordering technique. Another approach clusters vectors and merge groups of deterministic test cubes despite some degree of incompatibility [8]. A disadvantage of data compression for test is need of an ATE and the low diagnosis capability [9] [23]. Many works use built-in self-test (BIST) solutions [12][21-22] where either pseudo-random test patterns (PRTPs) and pre-computed test patterns (or deterministic test patterns) are stored in system memory. In terms of TAT and fault coverage, deterministic test patterns tend to be more effective than PRTPs [9]. Because of PRTPs cannot diagnosis hard-to-detect and random-resistant faults.

Some techniques use compaction algorithms [13] and decompression to minimize TAT [15]. Furthermore hybrid BIST methods have been used to decrease TAT [14][16][17]. The work presented in [18] reduced test application time with sharing scan chain and a combined method uses test data compression and test sharing to find the best test for each core. These shared chain algorithms are dependent on the number of ATE channels. In [20], the authors proposed a merging algorithm that merges test sets of different cores and broadcast shared tests to all CUTs, which will minimize the number of tests. But this algorithm has hardware redundancy. The work in [19] presents a highly accurate probabilistic method to generate fewer counterexamples to aid design debugging.

In this paper, we propose an improved scheme for generating deterministic tests. A simple algorithm is presented to cluster cores with some similar features. These clustered cores will be tested concurrently. Therefore, the TAT will be decreased significantly.

The rest of this paper is structured as follows. Section II is devoted to discuss SoC test architecture. A motivating example for illustrating main idea is presented in Section III. In Section IV an algorithm for core clustering is proposed. The results obtained by the proposed method are drawn in Section V. Finally Section VI concludes our work.

## 2. SoC Test Architecture

In this section, SoC test architecture will be described. Fig. 1 shows an overall view of a SoC architecture. This SoC consists of six cores, test memory and test controller which are connected together via a functional bus. Each core identified by its core number ($C_i$). Deterministic test vectors are in the test memory. In the heterogeneous SoC, deterministic test vectors are generated for each core independently and then applied to them. How much the cores in the SoC are increasing and SoC gets larger, the number of deterministic test vectors will be increased. Therefore the test application time will be large and unreasonable.

Because of using bus topology, at the same time, we can transfer one test vector to a core and the test time would be long.

Each core in the SoC has its own features, e.g., the number of inputs, outputs, gates, faults, etc. It's possible to generate shared deterministic vectors for the cores those have some features in common. Then these shared tests could be applied simultaneously to the cores. In the next section, this idea will be illustrated with architecture shown in Fig. 1. Then, an algorithm to select clusters for shared test generation will be presented.

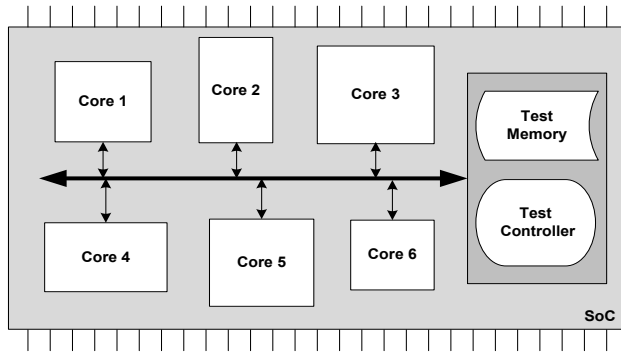

Fig. 1. SoC test architecture

## 3. Motivating Example

In this section, the basic idea will be explained with an example. Consider Fig. 1, the cores are ISCAS'89 benchmarks. With using ATALANTA, deterministic tests will be generated. Features associated with each core and the number of generated deterministic tests is drawn in Table 2.

PIs shows the number of primary inputs and POs shows the number of primary output of the circuit. FFs column is the number of flip-flops which are corresponded to the number of pseudo primary inputs (PPIs) and pseudo primary outputs (PPOs) of a core.

Ud_faults, FC and Det_tests mean the number of undetected faults, fault coverage and the number of deterministic tests (generated by ATALANTA) respectively.

Finally, Clk_cycle shows the number of clock cycles for applying generated patterns to each core.

Figure 2 shows the calculation of TAT for the SoC example. As we can see, TAT is the addition of test clock cycles for each core in the SoC.

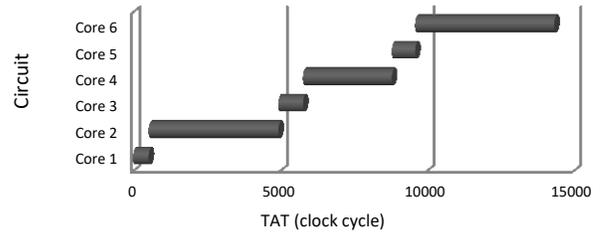

Fig. 2. TAT calculation of the SoC shown in Fig. 1

In Table 2, Core 1, 3, and 5 are similar in the number of inputs (PIs and FFs as PPIs). As well as, the number of gates for these three circuits is as close as possible. We can put these three cores into a cluster and generate deterministic tests for it. Core 2 and Core 6 are similar in the number of inputs and as close as possible in the number of gates. So we can put them into a cluster, too. Finally, Core 4 is put in a cluster independently.

Figure 3 shows the SoC after clustering the cores. As we can see, three clusters are created.

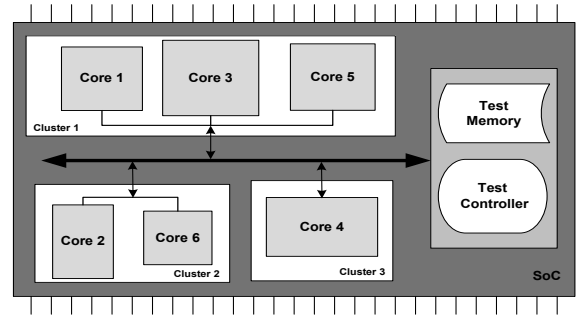

Fig. 3. Clustering the cores for the SoC example

In Table 2, the results for the number of generated deterministic tests for clusters are explained. Imp shows the percentage of the improvement in the number of test clock cycles in proportion to independent cores test generation.

CI shows the number of inputs for a cluster and has been determined by the maximum number of inputs of the SoC's cores.

The overall improvement of the SoC is obvious in Fig. 2 and Fig. 4 and it is calculated by Equation 1.

$$\text{Total\_Imp} = \frac{\sum \text{Core\_clk\_cycle} - \text{Cluster\_clk\_cycle}}{\sum \text{Core\_clk\_cycle}} \quad (1)$$

In Equation 1, core_clk_cycle is the number of clock cycles for applying generated pre-computed vectors for a core (before clustering). Cluster_clk_cycle shows the number of clock cycles for applying shared test vectors generated for a cluster. For the above example, the total improvement for TAT clock cycles is 39%.

Table 1. Generated tests of the clusters

| Cluster | Cores in cluster | #CI | #Det_tests | #Clk_cycle | Imp. (%) |
|---|---|---|---|---|---|
| 1 | 1,3,5 | 24 | 41 | 984 | 54.44 |
| 2 | 2,6 | 32 | 148 | 4736 | 48.25 |
| 3 | 4 | 54 | 56 | 3024 | 0 |

Table 2. Features for SoC Cores in Fig. 1

| Core number | Circuit name | #PIs | #POs | #FFs | #Gates | #Faults | #Ud_faults | FC(%) | #Det_tests | #Clk_cycle |
|---|---|---|---|---|---|---|---|---|---|---|
| Core 1 | S344 | 9 | 11 | 15 | 101 | 342 | 0 | 100 | 22 | 528 |
| Core 2 | S1196 | 14 | 14 | 18 | 388 | 1242 | 0 | 100 | 138 | 4416 |
| Core 3 | S382 | 3 | 6 | 21 | 99 | 399 | 0 | 100 | 35 | 840 |
| Core 4 | S713 | 35 | 23 | 19 | 139 | 581 | 38 | 93 | 56 | 3024 |
| Core 5 | S444 | 3 | 6 | 21 | 119 | 474 | 14 | 97 | 33 | 792 |
| Core 6 | S1238 | 14 | 14 | 18 | 428 | 1355 | 66 | 95 | 148 | 4736 |

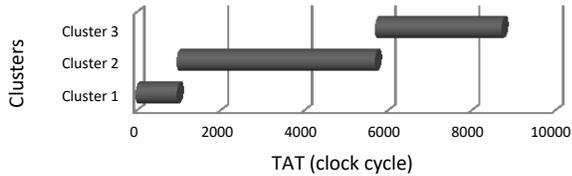

Fig. 4. TAT calculation of the SoC shown in Fig. 3

Fig. 4 shows the TAT of the SoC example, after clustering the cores. It is figured out from Fig. 2 and Fig. 4 that the number of test clock cycles after clustering has significantly reduced.

## 4. Clustering Algorithm

This section presents an algorithm for our proposed test generation. The features that we can gather some cores in a specific cluster are vary, i.e., the number of inputs including primary inputs (PIs), pseudo primary inputs (PPIs), the number of outputs including primary outputs (POs) and pseudo primary outputs (PPOs), the number of gates and the number of faults.

The following definitions are used to present clustering algorithm.

*Definition 1. Selecting cores for a cluster with as close as possible number of inputs and gates cause more reduction in the TAT.*

**Reasoning.** Suppose core 1 with 50 inputs and 80 deterministic test vectors. TAT of core 1 is 50×80=4000 clock cycles. Core 2 has 5 inputs and 14 deterministic patterns and TAT for core 2 is 5×14=70 clock cycles. Core 3 has 48 inputs and 72 deterministic tests and then, TAT for core 3 is 48×72=3456 clock cycles. Consider these three cores are similar in the number of gates. If we assume core 1 and core2 in a cluster and the number of common deterministic tests is obtained 84, the TAT for applying these vectors to their cluster (core 1 and core 2) is 84×(max number of inputs)= 84×50=4200 clock cycles. Since core 2 should wait 50 clock cycles for applying each deterministic vector instead of 5 clock cycles, the number of test clock cycles becomes worse.

Consider core 1 and core 3 in a cluster and the number of shared deterministic tests is obtained 90, the TAT for applying these vectors to their cluster (core 1 and core 3) is 90×(max number of inputs)= 90×50=4500 clock cycles. By using Equation 1, the improvement of the number of test clock cycle is 40%.

Consequently, selecting cores for a cluster with as close as possible number of inputs and gates are better and these similar features cause a greater reduction in TAT.

*Definition 2: If we cluster 2 cores with different number of inputs, the overall TAT will be increased.*

**Reasoning.** If two cores with different number of inputs are selected for a cluster, test application time will be increased against each core is tested independently. For example, in Table 3, the number of inputs for S1423 circuit is 91 and the number of test clock cycles is 5824. The number of inputs for S1488 circuit is 14 and the number of test clock cycles is 1764. By clustering these two cores and generate deterministic tests for this cluster, the number of deterministic tests will be decrease in proportion to independent test generation. But the number of clock cycle for testing this cluster will be increased dramatically.

Table 3. Clustering with different number of inputs

| Circuit name | #inputs | #Gates | #Det_tests | #Clk_cycle |
|---|---|---|---|---|
| S1423 | 91 | 490 | 64 | 5824 |
| S1488 | 14 | 550 | 126 | 1764 |
| S1423 & S1488 | 91 | 1040 | 162 | 14742 |

The following algorithm presents a simple approach for selecting cores and clustering. First, some definition should be described.

- $M_{SoC}$ is the set of clusters.
- $C_i$ shows the number of each core.
- $N_I$ is the number of inputs of the circuit. It is the addition of PIs and PPIs (number of FFs)
- $N_G$ shows the number of gates for each core.

First, $M_{SoC}$ is empty that shown in Line 1. For each core in the SoC, if there are some cores that they are similar in the number of inputs and gates (Line 2, 3), the generated deterministic tests for the clustered cores are less than independent cores. So, the number of test clock cycles will be decreased. Then we can put these cores into a cluster (Line 4). In the next step, there are some cores that they are similar in the number of inputs, but they are different in the number of gates. We choose these cores at the second priority of the clustering (Line 5, 6).

Algorithm 1. Core clustering

**Inputs**:
1) $C_i$ (Core Number), i = 0, ..., n
2) $N_I$ (Number of inputs, i.e., PIs and PPIs)
3) $N_G$ (Number of gates)

**Output**:
1) All $N_i$ sets ($M_{SoC}$)

**Finding $M_{SoC}$** ($C_i, N_I, N_G$)
1  $M_{SoC} = \{\};$
2  **FOR** $C_{i=1,...,n}$
3    **IF**($N_I(C_i) \approx N_I(C_j)$ and $N_G(C_i) \approx N_G(C_j))$ $(j = 1,...,n$ and $j \neq i)$
4        Include $C_i$, and all $C_j$ to in $N_i$;
5    **ELSEIF**($N_I(C_i) \approx N_I(C_j)$ and $N_G(C_i) ! \approx N_G(C_j)$ $(j =$
6          $1,...,n$ and $j \neq i)$
7        Include $C_i$, and all $C_j$ to in $N_i$;
8    **ELSE**
9        Consider $C_i$ as independent clusters;
10   **ENDIF**
11 **ENDFOR**
  **END**

## 5. Experimental Results

The experimental results for ISCAS'89 are shown in Table 4 and Table 5. Table 4 shows some features for each core, e.g. the number of inputs, the number of outputs, the number of faults and the number of generated deterministic test vectors. Cores 1, 2, 3, 4 and 5 are similar in the number of inputs and as close as possible in the number of gates.

If a cluster becomes very big, we should consider power consumption in the test mode and it does not exceed to maximum power.

In Table 5, the clusters are shown. As we can see, in the first four clusters, in each row, we add one core with the same feature to the cluster and the improvement for TAT increases. The Imp calculates from Equation 1 in Section III.

## 6. Summary and Conclusion

As the technology advances, testing of SoCs has remained a challenging problem. In addition, systems tend to become larger and more complex in each technology generation. This situation has made test application time dramatically high. In this paper an improved scheme for pre-computed patterns in core-based SoCs is presented that attempts to reduce TAT by clustering cores with the same features. This scheme reduces the number of required deterministic test data significantly.

Furthermore, reduction in total number of pre-computed test data causes the cost of communication decreases significantly specially in large systems like large SoCs and MPSoCs.

Table 4. The features for some cores

| Core number | Circuit name | #PIs | #Pos | #FFs | #Gates | #Faults | #Ud_faults | FC(%) | #Det_tests | #Clk_cycle |
|---|---|---|---|---|---|---|---|---|---|---|
| *1* | S344 | 9 | 11 | 15 | 101 | 342 | 0 | 100 | 22 | 528 |
| *2* | S349 | 9 | 11 | 15 | 104 | 350 | 2 | 99 | 21 | 504 |
| *3* | S382 | 3 | 6 | 21 | 99 | 399 | 0 | 100 | 35 | 840 |
| *4* | S400 | 3 | 6 | 21 | 106 | 424 | 6 | 99 | 35 | 840 |
| *5* | S444 | 3 | 6 | 21 | 119 | 474 | 14 | 97 | 33 | 792 |
| *6* | S713 | 35 | 23 | 19 | 139 | 581 | 38 | 93 | 56 | 3024 |
| *7* | S820 | 18 | 19 | 33 | 256 | 850 | 0 | 100 | 112 | 5712 |
| *8* | S838 | 34 | 1 | 32 | 288 | 857 | 0 | 100 | 99 | 6534 |
| *9* | S953 | 16 | 23 | 29 | 311 | 1079 | 0 | 100 | 91 | 4095 |
| *10* | S1196 | 14 | 14 | 18 | 388 | 1242 | 0 | 100 | 138 | 4416 |
| *11* | S1238 | 14 | 14 | 18 | 428 | 1355 | 66 | 95 | 148 | 4736 |
| *12* | S1423 | 17 | 5 | 74 | 490 | 1515 | 14 | 99 | 64 | 5824 |
| *13* | S1488 | 8 | 19 | 6 | 550 | 1486 | 0 | 100 | 126 | 1764 |
| *14* | S1494 | 8 | 19 | 6 | 558 | 1506 | 12 | 99 | 129 | 1806 |
| *15* | S5378 | 35 | 49 | 179 | 1004 | 4603 | 40 | 99 | 254 | 54356 |
| *16* | S9234 | 19 | 22 | 228 | 2027 | 6927 | 404 | 94 | 380 | 93860 |

Table 5. Test generation for clusters

| Clusters | #CI | #Gates | #Det. tests | FC (%) | #Clk cycle | Imp. (%) |
|---|---|---|---|---|---|---|
| 1,2 | 24 | 205 | 23 | 99.7 | 552 | 46.51 |
| 1,2,3 | 24 | 304 | 39 | 99.8 | 936 | 50 |
| 1,2,3,4 | 24 | 410 | 40 | 99.4 | 960 | 64.60 |
| 1,2,3,4,5 | 24 | 529 | 51 | 98.9 | 1224 | 65.06 |
| 1,2,3,7 | 51 | 560 | 126 | 99.9 | 6426 | 15.26 |
| 1,2,3,10 | 32 | 692 | 157 | 99.9 | 5024 | 20.10 |
| 10,11 | 32 | 816 | 148 | 97.4 | 4736 | 48.25 |
| 13,14 | 14 | 1108 | 130 | 99.6 | 1820 | 49.01 |
| 15,16 | 247 | 3031 | 562 | 95.8 | 138814 | 6.34 |